\documentclass[ sigconf]{acmart}
\AtBeginDocument{
  \providecommand\BibTeX{{
    \normalfont B\kern-0.5em{\scshape i\kern-0.25em b}\kern-0.8em\TeX}}}
\usepackage[utf8]{inputenc}
\usepackage{color,soul}
\usepackage{mathtools}
\usepackage{algorithm,algpseudocode}
\usepackage[binary-units,per-mode=symbol,range-phrase=~--~]{siunitx}
\usepackage{flexisym}
\usepackage[export]{adjustbox}
\usepackage{footnote}
\usepackage{makecell}
\usepackage{xurl}
\usepackage{comment}
\usepackage{tikz}
\usepackage{caption}
\usepackage{subcaption}
\usepackage{graphicx}
\usepackage[inline,shortlabels]{enumitem}
\usepackage{multirow}
\usepackage{adjustbox}
\usepackage{lipsum}
\usepackage{balance}
\newcommand{\datacloud}{DataCloud}
\newcommand{\defpipe}{DEF-PIPE}
\newcommand{\simpipe}{SIM-PIPE}
\newcommand{\adapipe}{ADA-PIPE}
\newcommand{\deppipe}{DEP-PIPE}

\newcommand{\deadline}{\mathtt{DEADLINE}}
\newcommand{\CORE}{\mathtt{CORE}}
\newcommand{\CPU}{\mathtt{CPU}}
\newcommand{\MEM}{\mathtt{MEM}}

\newcommand{\SOURCE}{\mathcal{P}}

\newcommand{\DATA}{\mathtt{data}}

\newcommand{\sched}{\mathtt{sched}}

\newcommand{\mt}{\mathtt{MT}}
\newcommand{\mcr}{\mathtt{MCR}}
\newcommand{\weight}{\mathtt{w}}
\newcommand{\bias}{\mathtt{b}}
\newcommand{\REQ}{\mathtt{req}}

\begin{document}
\title{Comparison of Microservice Call Rate Predictions \\for Replication in the Cloud}
\author{
Narges Mehran$^{\star}$, Arman Haghighi$^{\dagger}$, Pedram Aminharati$^{\star}$, Nikolay Nikolov$^{\ddagger}$\\
Ahmet Soylu$^{\mathsection}$, Dumitru Roman$^{\ddagger\mathsection}$,
Radu Prodan$^{\star}$}

\affiliation{\institution{$^{\star}$Alpen-Adria-Universit{\"a}t Klagenfurt \country{Austria} }}
\affiliation{\institution{$^{\dagger}$Azad University, Science and Research Branch, Tehran \country{Iran}}}
\affiliation{\institution{$^{\ddagger}$SINTEF AS, Oslo \country{Norway}}}
\affiliation{\institution{$^{\mathsection}$OsloMet - Oslo Metropolitan University, Oslo \country{Norway}}}
\email{{narges.mehran, radu.prodan}@aau.at}\email{armanhku@gmail.com}\email{{nikolay.nikolov,dumitru.roman}@sintef.no}\email{ahmet.soylu@oslomet.no}



\begin{abstract}
Today, many users deploy their microservice-based applications with various interconnections on a cluster of Cloud machines, subject to stochastic changes due to dynamic user requirements. To address this problem, we compare three machine learning (ML) models for predicting the microservice call rates based on the microservice times and aiming at estimating the scalability requirements. We apply the linear regression (LR), multilayer perceptron (MLP), and gradient boosting regression (GBR) models on the Alibaba microservice traces. The prediction results reveal that the LR model reaches a lower training time than the GBR and MLP models. However, the GBR reduces the mean absolute error and the mean absolute percentage error compared to LR and MLP models. Moreover, the prediction results show that the required number of replicas for each microservice by the gradient boosting model is close to the actual test data without any prediction.
\end{abstract}
\maketitle
\textcolor{red}{\scriptsize 2023 ACM/IEEE.  Personal use of this material is permitted.  Permission from ACM/IEEE must be obtained for all other uses, in any current or future media, including reprinting/republishing this material for advertising or promotional purposes, creating new collective works, for resale or redistribution to servers or lists, or reuse of any copyrighted component of this work in other works.}

\section{Introduction} \label{sec:intro}
The recent shift towards the increasing number of microservice-based applications in the Cloud-native infrastructure brings new scheduling, deployment, and orchestration challenges~\cite{joseph2020intma}, such as scaling out overloaded microservices in response to increasing load.

\subsubsection*{Research problem} 
inspected in this work, extends our previous work~\cite{mehran2022matching}, where we explored microservice scheduling on provisioned resources. 
In~\cite{mehran2022matching}, we did not inspect the scalability requirements of the containerized microservices by prediction models considering different request arrival rates from end-users acting as producers~\cite{luo2021characterizing}. 
Traditional microservice scaling methods~\cite{arkian2021model,rzadca2020autopilot} focus on the resource or application processing metrics without predicting the stochastic changes in user requirements, such as dynamic request rates.   

\subsubsection*{Example} tabulated in Table~\ref{tabl:example} presents an example involving three producers calling three microservices deployed on three resources. In this scenario, the microservices experience varying \emph{call rates} initiated by the \emph{producers}. Every producer request leads to interactions with its corresponding microservice on the specific resource within a specific time. Typically, microservices with higher execution times necessitate horizontal scalability to accommodate the call rate. In other words, a direct correlation exists between the microservice time and call rate, motivating the need to explore prediction models addressing their horizontal scaling~\cite{horn2022multi}. Table~\ref{tabl:example} shows that during a \SI{2}{\second} execution, the microservices $m_0$, $m_1$, and $m_2$ receive the following number of calls:
\begin{equation*}
\begin{array}{ll}
m_0: & \SI{2}{\second}\cdot\SI{2}{calls\per\second}=\SI{4}{calls};\\ 
m_1: & \SI{2}{\second}\cdot\SI{2}{calls\per\second}=\SI{4}{calls};\\
m_2: & \SI{2}{\second}\cdot\SI{3}{calls\per\second}=\SI{6}{calls}.
\end{array}
\end{equation*}
However, at the end of the \SI{2}{\second} interval, the microservices $m_0$, $m_1$, and $m_2$ still respond to their third, second, and first calls.
To reduce the bottleneck on the Cloud infrastructure~\cite{nikolov2021conceptualization}, we need to scale the microservices based on the multiplication function between the correlated microservice time and call rate up to the following number of replicas:
\begin{equation*}
\begin{array}{ll}
m_0 \mathrm{~on~} r_0: & \SI{2}{calls\per\second}\cdot\SI{0.7}{\second\per call}=\num{1.4}{}\approx \num{2};\\
m_1 \mathrm{~on~} r_1: & \SI{2}{calls\per\second}\cdot\SI{1.5}{\second\per call}=\num{3};\\
m_2 \mathrm{~on~} r_2: & \SI{3}{calls\per\second}\cdot\SI{2}{\second\per call}=\num{6}.
\end{array}
\end{equation*}

\begin{table}[t]
\centering
\caption{Motivational example.}
\label{tabl:example}
\resizebox{\columnwidth}{!}{
\begin{tabular}{|@{ }c@{ }|@{ }c@{ }|@{ }c@{ }|@{ }c@{ }|}
     \hline
 \emph{Microservice} & \emph{Resource} & \emph{Microservice time (\si{\second\per call})} & \emph{Call rate (\si{calls\per\second})}\\\hline
   $m_0$  & $r_0$ & \num{0.7} &  \num{2}\\\hline
   $m_1$  & $r_1$ & \num{1.5} &  \num{2}\\\hline
   $m_2$  & $r_2$ & \num{2}   &  \num{3}\\\hline
\end{tabular}}
\end{table}

\subsubsection*{Method} proposed in this work, addresses the scalability problem through \emph{microservices call rate predictions} 
employing ML models involving two features:
\begin{itemize}
\item \emph{Microservice time} defining the processing time of each containerized microservice on the Cloud virtual machine;
\item \emph{Microservice call rate} defining the number of calls/requests invoking a microservice.
\end{itemize}
We apply ML models to predict microservices call rate based on the microservice time and estimate the number of microservice replicas to support stochastic changes due to the dynamic user requirements. 
Recently, there has been a growing interest in the applicability of deep learning models to tabular data~\cite{gorishniy2023revisiting, gorishniy2023embeddings}. However, tree-based machine learning (ML) models such as bagging (e.g., RandomForest) or boosting (e.g., XGBoost~\cite{chen2015xgboost}, gradient boosting tree, and gradient boosting regression) are among the popular learners for tabular data that outperform deep learning methods~\cite{grinsztajn2022tree}. Nevertheless, related work did not explore and evaluate the \emph{gradient boosting regression (GBR)} and \emph{multilayer perceptron (MLP)} learning methods for microservice call rate prediction. Therefore, we apply and compare the GBR, neural network-based MLP, and traditional linear regression (LR) models to estimate the number of replicas for each microservice.  


\subsubsection*{Contributions} comprise a comparative evaluation of the ML models on trace data collected from a real-word Alibaba Cloud cluster~\cite{luo2022depth} indicating that the GBR reaches a balance between the prediction errors, including the mean absolute error (MAE) and mean absolute percentage error (MAPE), and the training time compared to the MLP and LR methods.

\subsubsection*{Outline} The paper has eight sections. We survey the related work in Section~\ref{sec:related}. Section~\ref{sec:model} describes the application, resource, schedule, microservice time models, and the main objective, followed by the ML prediction models in Section~\ref{sec:predmodels}. Section~\ref{sec:arch} presents the 
architecture of the replication predictions. Section~\ref{sec:experimntdesign} describes the experimental design and evaluation, followed by the results presented in Section~\ref{sec:experimntresult}. Finally, Section~\ref{sec:conclusion} concludes the paper.

\section{Related Work} \label{sec:related}
This section reviews the state-of-the-art analysis of microservice traces, workload prediction, and autoscaling of microservices in the Cloud infrastructures.

\subsubsection*{Microservice prediction}
Luo et al.~\cite{luo2022power} designed a proactive workload scheduling method by adopting CPU and memory utilization to ensure service level agreements while scaling up the resources. The work in~\cite{rahman2019predicting} predicted the end-to-end latency between microservices in the Cloud based on the MLP, LR, and GBR models. Cheng et al.~\cite{cheng2017high} applied GBR for predicting the resource requirements to execute the user's workload. Rossi et al.~\cite{rossi2020geo} proposed a reinforcement learning scaling method based on the microservice time. \c{S}tefan et al.~\cite{eInformatica2022Art07} presented a deep learning-based workload prediction to autoscale microservices, highlighting the MLP model.

\subsubsection*{Alibaba microservice trace analysis}
Luo et al.~\cite{luo2022depth} explored the large-scale deployments of microservices based on their dependencies and the runtime execution times on the Alibaba Cloud clusters and showed that service response time tightly relies on the call graph topology among microservices that impacts the runtime performance. He et al.~\cite{hegraphgru2023} proposed a graph attention network-based method to predict the resource usages based on the topological relationships among the Cloud physical machines and validated this method through the Alibaba microservice dataset.

\subsubsection*{Autoscaling}
Arkian et al.~\cite{arkian2021model} presented a geo-distributed auto-scaling model for the Apache Flink framework to sustain the throughput among resources, optimizing the network latency and resource utilization. 
Autopilot~\cite{rzadca2020autopilot} proposed a method to scale in/out the number of replicas from each service in a time interval (e.g., \SI{5}{\minute}) based on the $\CPU$ usage and the average required utilization of the microservices.

\subsubsection*{Research gap} Related methods designed the microservice prediction models based on the completion time or the cost. We extend these methods by researching microservice call rate and replica prediction based on the microservice time using LR, GBR, and MLP machine learning models.

\section{Data Processing Model} \label{sec:model}
This section presents the formal model underneath our work.

\subsubsection*{Data processing streams} \mbox{$\mathcal{S} = \left(\mathcal{M},\mathcal{M}_{\SOURCE}, \mathcal{D}\right)$} consist of:
\paragraph{Microservices}  representing  a set of independent tasks $\mathcal{M} =\left\{m_i\ |\ 0 \leq i < \mathcal{N}_{\mathcal{M}}\right\}$.
\paragraph{Producer} $\SOURCE\in \mathcal{M}_{\SOURCE}$ generating data at the rate $\mcr_{\SOURCE{}i}$ that requires further processing by a microservice $m_i$. 
\paragraph{Dataflow} $\DATA_{\SOURCE{}i}$ streaming from a producer $\SOURCE\in \mathcal{M}_{\SOURCE}$ to a microservice $m_i \in \mathcal{M}$: \mbox{$\mathcal{D} = \left\{ \left(\SOURCE, m_i, \DATA_{\SOURCE{}i}\right) | \left(\SOURCE, m_i\right) \in \mathcal{M}_{\SOURCE} \times \mathcal{M}\right\}$}. 
\paragraph{Resource requirements} $\REQ\left(m_i\right)$ for proper processing of a dataflow $\DATA_{\SOURCE{}i}$ by a microservice $m_i$ is a pair representing the minimum number of cores $\CORE\left(m_i\right)$, memory $\MEM\left(m_i\right)$ size (in \si{\mega\byte}), and deadline for execution (in \si{\second})~\cite{samani2023incremental}:
\begin{equation*}
\REQ\left(m_i\right) = \left(\CORE\left(m_i\right), \MEM\left(m_i\right), \deadline\left(m_i\right)\right).
\end{equation*}
\paragraph{Minimum processing load} $\CPU\left(m_i\right)$ is the (million) number of instructions $(\si{MI})$) of dataflow $\DATA_{\SOURCE{}i}$ processed by a microservice $m_i$.

\subsubsection*{Resources} $\mathcal{R} = \left\{r_j | 0 \leq j < \mathcal{N}_{\mathcal{R}}\right\}$ represent a set of $\mathcal{N}_{\mathcal{R}}$ Cloud virtual machines. We define a resource \mbox{$r_j = \left(\CORE_j, \MEM_j\right)$} as a vector representing its available processing core $\CORE_j$ and memory $\MEM_j$ size (in \si{\giga\byte}), depending on its utilization. Every device has an available processing speed denoted as $\CPU_j$ (in \si{MI} per second).

\subsubsection*{Schedule} of microservice $m_i$ is a mapping on a resource $r_j = \sched(m_i)$ that satisfies its processing and memory requirements: $\CORE\left(m_i\right)\leq\CORE_j \ \land\ \MEM\left(m_i\right) \leq \MEM_j\ \land \mt_{i,j}\leq\deadline\left(m_i\right)$, where $\mt{i,j}$ is the microservice time defined in the next paragraph.

\subsubsection*{Microservice time} $\mt\left(m_i,r_j\right)$ or $\mt_{i,j}$ of $m_i$ on a resource $r_j=\sched\left(m_i\right)$ is the ratio between its computational workload $\CPU\left(m_i\right)$ (in \si{MI}) and the processing speed $\CPU_j$ (in \si{MI} per second)~\cite{mehran2022matching}:
\begin{equation*}
\mt_{i,j} = \frac{\CPU\left(m_i\right)}{\CPU_j}.
\label{eq:microtime}
\end{equation*}

\subsubsection*{Objective} \label{sec:obj} is to estimate the number of \emph{replicas} $\mathcal{L}_{ij}$ for horizontally scaling a microservice $m_i$ based on the producer call rate $\mcr_{\mathcal{P}i}$ and its microservice time $\mt_{i,j}$ on a resource $r_j$: $\mathcal{L}_i = \mcr_{\mathcal{P}i} \cdot \mt_{i,j}$.

\section{Prediction Models} \label{sec:predmodels}
This section summarizes the ML models used in this paper for predicting microservice call rates based on their service times, further used to decide their replicas. 

\subsubsection*{Linear regression}  \label{sec:predmodels:lr} defines a relation between the microservice time $\mt_{i,j}$ (as the input feature to the model) and the microservice call rate $\mcr_{\SOURCE{}i}$ (as the output feature of the model), where $r_j = \sched\left(m_i\right)$. Thereafter, we model a linear relation between the predicted microservice call rate $\mcr^{\prime}_{\SOURCE{}i}$ and the actual microservice time $\mt_{i,j}$:
\begin{equation*}
    \mcr^{\prime}_{\SOURCE{}i} = 
    \mt_{i,j}\cdot \weight_{\SOURCE{}i}+\bias_{\SOURCE{}i},
\end{equation*}
where $\weight_{\SOURCE{}i}$ and $\bias_{\SOURCE{}i}$ denote the \emph{weight} and \emph{bias} of the LR model,
learned to fit a linear relation between the predicted microservice call rate $\mcr^{\prime}_{\SOURCE{}i}$ and the microservice time $\mt_{i,j}$.
The microservice weights form a set: \[\weight =\{\weight_{\SOURCE{}i}|\SOURCE\in \mathcal{M}_{\SOURCE}\ \land\ 0\le i<\mathcal{N}_{\mathcal{M}}\},\]
calculated 
by minimizing the sum of the squared differences between the predicted microservice call rate $\mcr{^\prime}_{\SOURCE{}i}$ and the weighted microservice time $\mt_{i,j}\cdot \weight_{\SOURCE{}i}$~\cite{scikit-lr}:
\begin{equation*}
\min_{\weight}\sum_{\weight_{\SOURCE{}i}\in\weight}\left(\mcr^{\prime}_{\SOURCE{}i} -\mt_{i,j}\cdot \weight_{\SOURCE{}i}\right)^2.
\end{equation*}
Moreover, the biases have independent and identical normal distributions with mean zero and constant variance~\cite{lr.mathwork}.

\subsubsection*{Multilayer perceptron} 
\begin{figure}[!t]
    \centering    \includegraphics[width=0.9\columnwidth]{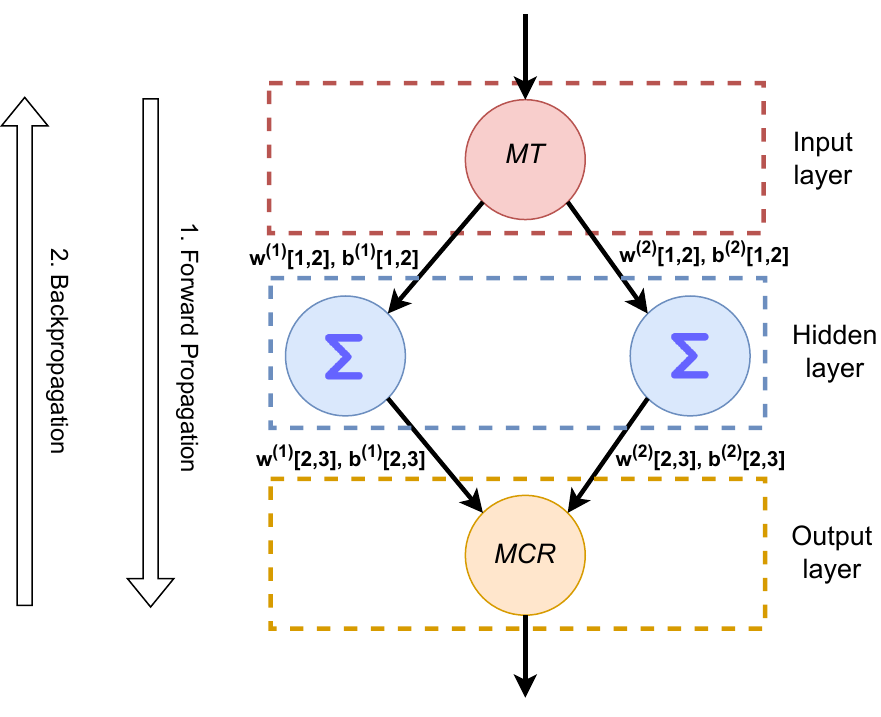}
    \caption{MLP neural network architecture.}
    \label{fig:mlp}
\end{figure}

belongs to the category of the feedforward artificial neural network, comprising a minimum of three layers of neurons~\cite{keeni1999estimation}: an input layer, one or more hidden layers, and an output layer (see Figure~\ref{fig:mlp}). This algorithm combines inputs with initial weights in a weighted sum and subsequently passes through an activation function and mirrors the process observed in the perceptron~\cite{nn.linear}. This model propagates backward through the ML layers and iteratively trains the partial output of the loss function to update the model parameters (see Figure~\ref{fig:mlp}):
\begin{multline*}
    \mcr^{\prime}_{\SOURCE{}i} = \sum\limits_{h=1}^{ \mathcal{N}_{\mathcal{N}}}\Bigg(\big(\mt_{i,j}\cdot \weight_{\SOURCE{}i}^{(1)}[x,h] +\bias_{\SOURCE{}i}^{(1)}[x,h]\big)\cdot\\\cdot\weight_{\SOURCE{}i}^{(2)}[h,y]+ \bias_{\SOURCE{}i}^{(2)}[h,y]\Bigg),
\end{multline*}
where $\weight_{\SOURCE{}i}$ and $\bias_{\SOURCE{}i}$ denote the learnable weight and bias of the linear MLP model and $\mathcal{N}_{\mathcal{N}}$ defines the number of neurons in a hidden layer~\cite{keeni1999estimation}. This paper defines a single-neuron input layer $x=1$ and a corresponding output layer $y=3$.

\subsubsection*{Gradient boosting regression} estimates and constructs an additive model in a forward stage-wise manner. GBR~\cite{natekin2013gradient} can ensemble multiple prediction models (e.g., regression trees) to create a more accurate model~\cite{zhong2022machine}.
\begin{equation*}
\mcr^{\prime}_{\SOURCE{}i} = F_{\mathcal{E}}(\mt_{i,j}) = \sum_{e=1}^{\mathcal{E}} h_e(\mt_{i,j}),
\end{equation*}
where the $h_e$ is a boosting estimator and $\mathcal{E}$ is a constant corresponding to the number of estimators~\cite{ensemble.methods} used by the fixed-size decision tree regressors.

\section{Architecture Design} \label{sec:arch}
We present in this section the architecture design of our method implemented in the $\datacloud$ toolbox~\cite{nikolov2023container}.

\subsection{$\datacloud$}
We designed the architecture of our method in the context of the $\datacloud$~\cite{roman2022big} project supporting the lifecycle of microservices-based applications processing streams and batches of data on the computing continuum through the interaction of four tools.


\paragraph{$\defpipe$} defines the application services and structure from the user input using a domain-specific language model to define the microservice requirements~\cite{tahmasebi2022dataclouddsl};
\paragraph{$\simpipe$} simulates the dataflow execution based on the microservice's processing speed and memory size requirements before large-scale deployment~\cite{thomas2022sim};
\paragraph{$\adapipe$} receives the microservices, explores their requirements, such as processing and memory size, predicts the number of replicas for each microservice based on its resource requirements, adapts the execution, and sends to $\deppipe$ for the deployment;
\paragraph{$\deppipe$} deploys the dataflow processing microservices on the computing resources based on $\adapipe$ schedules and manages their execution on multiple Kubernetes clusters at the user's location or in the Cloud~\cite{simonet2022toward}.

\subsection{$\adapipe$}
\begin{figure}[!t]
    \centering
    \includegraphics[width=\columnwidth]{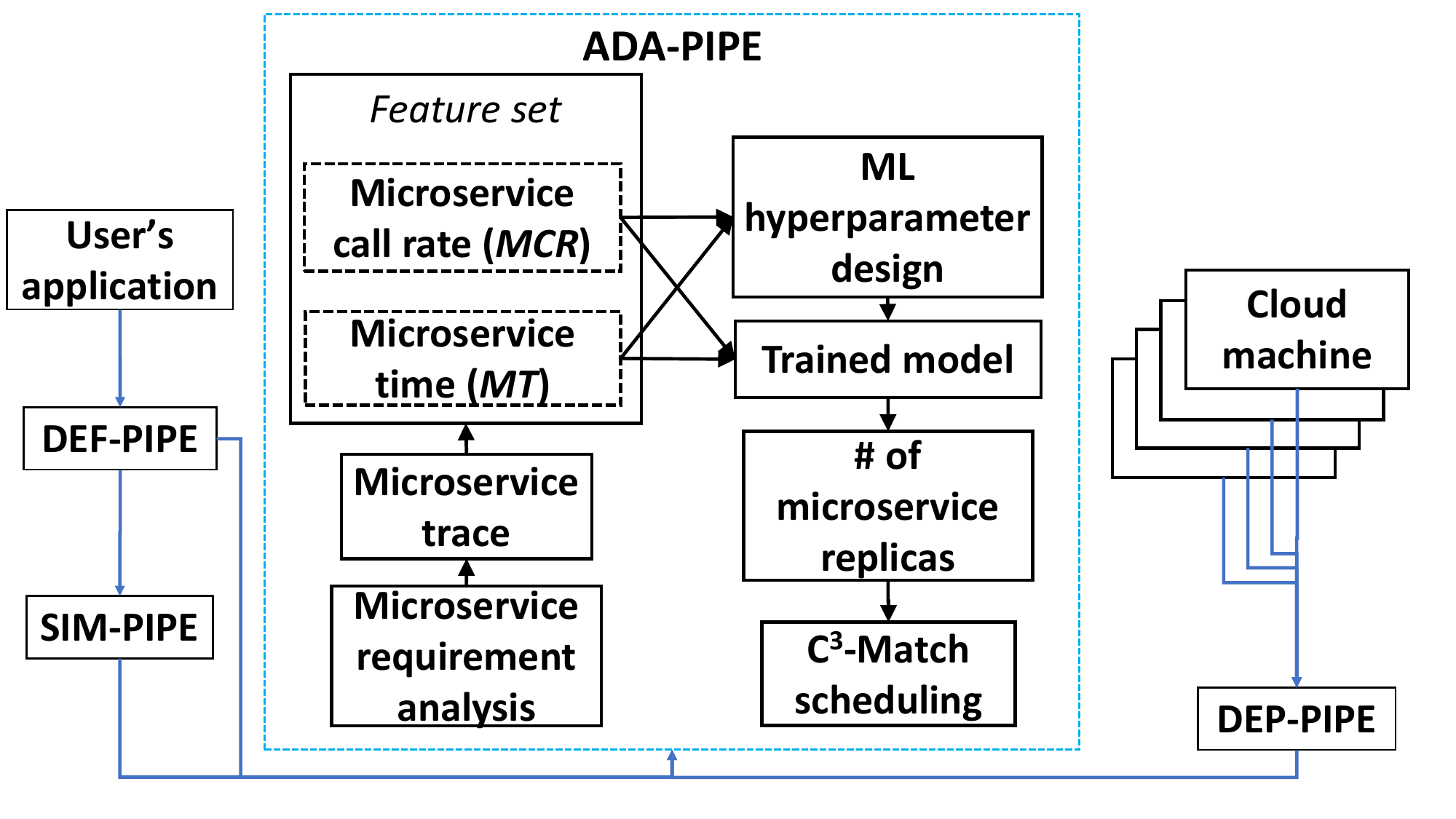}
    \caption{$\adapipe$ architecture.}
    \label{fig:arch}
\end{figure}

Figure~\ref{fig:arch} illustrates the $\adapipe$'s component with its replica prediction, consisting of seven components.

\paragraph{Microservice requirement analysis} receives resource needs $\REQ\left(m_i\right)$ of the user application microservices to update the trace. 
Moreover, it includes the simulation information of the microservice times provided by $\simpipe$ to predict the microservices to the trace;

\paragraph{Microservice trace} consists of rows with the timestamp, microservice name, microservice container instance identifier, and the collected metrics (i.e., $\mcr_{\SOURCE{}i}$, $\mt_{i,j}$)~\cite{luo2021characterizing};

\paragraph{Feature set} receives the dataset and extracts the microservice call rate $\mcr_{\SOURCE{}i}$ for a corresponding microservice time $\mt_{i,j}$. The ML models learn to fit the $\mt$ as the input to the $\mcr_{\SOURCE{}i}$ as the output;

\paragraph{ML hyperparameter design} component receives the feature set consisting of $\mcr_{\SOURCE{}i}$ and $\mt_{i,j}$ for fine-tuning and optimizing the ML models. It utilizes an exhaustive search to configure and tune the hyperparameters;

\paragraph{Prediction model} forecasts the microservice call rate based on the microservice time by utilizing the ML prediction models, including LR, MLP, and GBR (see Section~\ref{sec:predmodels});

\paragraph{Replica} component estimates the required instances to scale out from each microservice based on the multiplication between its predicted call rate $\mcr{^\prime}_{\SOURCE{}i}$ and time $\mt_{i,j}$; 

\paragraph{Orchestration} manages the microservices on the Cloud virtual machines by utilizing the \texttt{Kubernetes} replica scaling\footnote{\url{https://github.com/SiNa88/HPA}}  based on the predicted microservice call rates and decisions taken by the integrated scheduler~\cite{mehran2022matching}.

\section{Experimental Design} \label{sec:experimntdesign} 
This section presents our experimental design for the dataset preparation, testbed, and tuning of hyperparameters. 

\subsection{Dataset preparation}
We validated our method using simulation based on an Alibaba microservice dataset\footnote{\url{https://shorturl.at/fjsSU}} 
available in a public repository\footnote{\url{https://zenodo.org/record/8310376}}. 
The dataset contains dataflows with various communication paradigms among over \num{1300} microservices running on more than \num{90000} containers for twelve hours, recorded in a time interval of \SI{30}{\second}~\cite{luo2021characterizing}. We selected \num{180000} rows of the dataset, denoting the microservice times \mbox{$\mt_{i,j}$= \SIrange{0.01}{5859}{\milli\second\per call}}, and the microservice call rates \mbox{$\mcr_{\SOURCE{}i}$ = \SIrange{0.025}{4874}{calls\per\second}}. 

\subsection{Testbed design}
We implemented ML algorithms in \texttt{Python 3.9} using \texttt{scikit-learn} API~\cite{scikitlearn_api}.
Afterward, we compared the runtime performance of the algorithms on two machines:
\begin{itemize}
\item \emph{Google CoLaboratory (CoLab)} with NVIDIA$^\circledR$ Tesla$^{(TM)}$ T4 GPU accelerator and \SI{16}{\giga\byte} of memory\footnote{\url{https://colab.research.google.com/}};
\item \emph{Personal device} with an \num{8}-core Intel$^\circledR$ Core$^{(TM)}$ \mbox{i7-7600U} processor and \SI{16}{\giga\byte} of memory.
\end{itemize}

\subsection{ML hyperparameter design}
This section presents the learning procedure of fine-tuning and optimization of the hyperparameters of the GBR and MLP models, summarized in Table~\ref{tab:hypset}, based on three steps: exhaustive search, hyperparameter tuning, and hyperparameter configuration. However, we rely on the default settings of ordinary least squares optimization for the LR model~\cite{scikit-lr}.

\subsubsection{Gradient boosting regressor} 
\begin{figure}[!t]
    \centering
    \subfloat[GBR loss]{\includegraphics[width=.49\columnwidth]{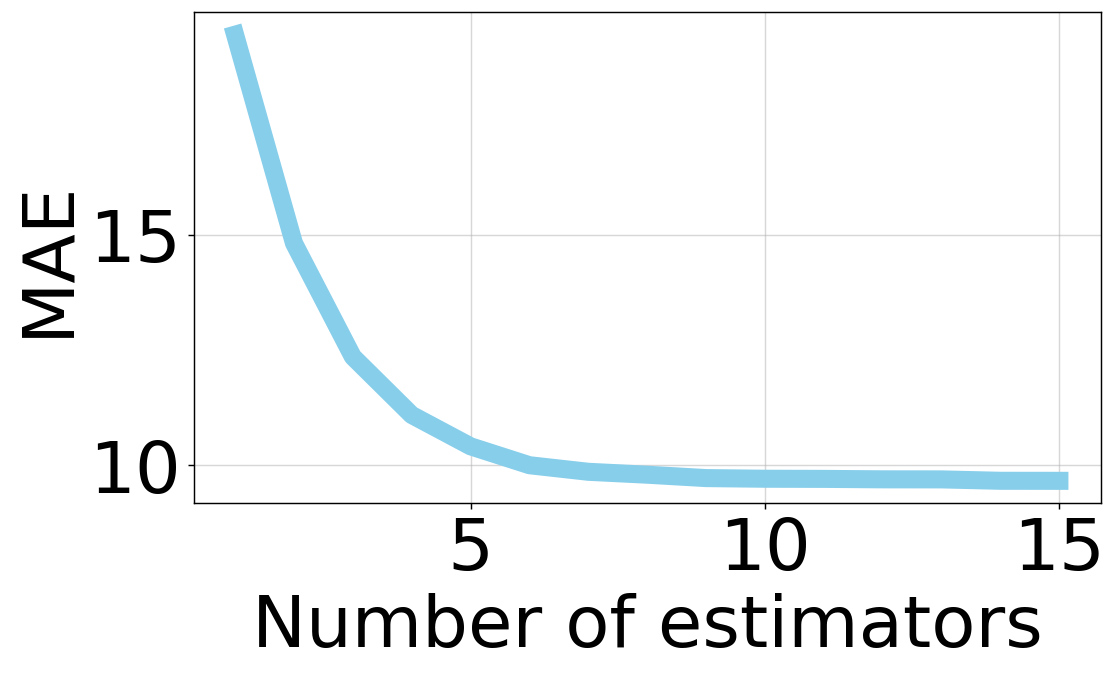}\label{fig:gbr-val}}
    \hfill
    \subfloat[MLP loss]{\includegraphics[width=.49\columnwidth]{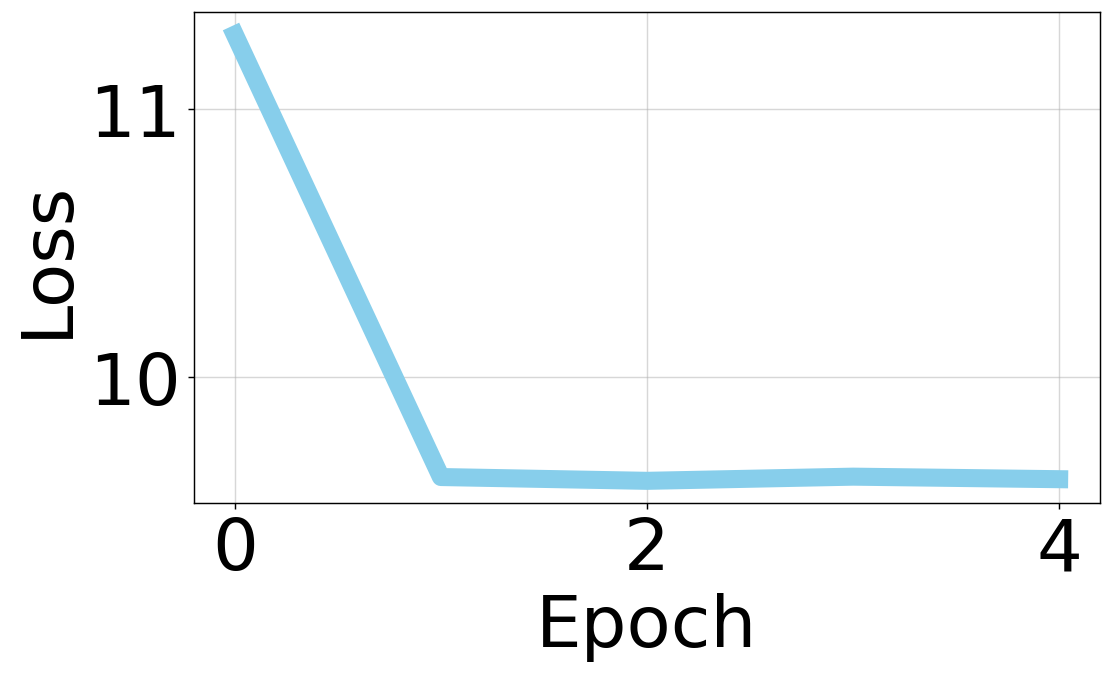}\label{fig:mlp-val}}
    \caption{GBR and MLP losses.}
\end{figure}

uses a learning curve to evaluate the changes in the training loss for different iterations based on the number of evaluators and the learning rate  (see Figure~\ref{fig:gbr-val}).

\paragraph{Exhaustive search} uses the \texttt{GridSearchCV} library of the ML toolkit \texttt{scikit-learn} and sets the number of estimators to \num{300} and learning rate to \num{0.02}, which results in overfitting the training data. 

\paragraph{Hyperparameter tuning} modifies the number of estimators in the range of \numrange{10}{170} and the learning rate in the range of \numrange{0.02}{0.4} to converge to a stability point with a faster training time.

\paragraph{Hyperparameter configuration}
sets the number of estimators to \num{15} and the learning to rate to \num{0.4} with an improved training score without overfitting and reduced training time. 
Figure~\ref{fig:gbr-val} shows that, during the training loop, the model tunes each gradient tree or estimator to the previous tree model's error until it reaches the maximum number of estimators set.

\subsubsection{Multilayer perceptron} uses the three-step learning curve evaluating the ML model’s performance through the changes in the training loss with different training iterations, number of layers, and number of neurons in the network (see Figure~\ref{fig:mlp-val}).

\paragraph{Exhaustive search} uses the \texttt{PyTorch} library and sets the number of  hidden layers to \num{3}, the number of neurons to \num{100}, and the learning rate to \num{0.4} in \num{50} epochs, overfitting the training data.
\paragraph{Hyperparameter tuning} decreases the learning rate to \num{0.0005} to comprehend more about the training procedure. Afterward, we reduce the complexity of the model by lowering the number of neurons and hidden layers (see Figure~\ref{fig:mlp}) in each epoch because of the single input of $\mt_{i,j}$ in the dataset.

\paragraph{Hyperparameter configuration} of the MLP model with two epochs, one hidden layer of two neurons, and a learning rate of \num{0.003} predicts the $\mcr_{\SOURCE{}i}$ without overfitting, as shown in Figure~\ref{fig:mlp-val}.

\begin{table}[t]
\centering
\caption{ML hyperparameter design.}
\label{tab:hypset}
\resizebox{\columnwidth}{!}{
\begin{tabular}{|@{ }c@{ }|@{ }c@{ }|@{ }c@{ }|@{ }c@{ }|}
\hline
\textit{Model} & \textit{Hyperparameter} &\textit{Description} & \textit{Value}           \\\hline\hline
\multirow{4}{*}{MLP} & Hidden layers        & Number of hidden layers                                                             & 1               \\
\cline{2-4}
                     & Optimizer          & Optimization algorithm                                                     & Adam            \\\cline{2-4}
                     & Learning rate      & Learning step size at each iteration                                                    & \num{0.003}           \\\cline{2-4}
                     & Loss        & Error quantification between $\mcr^{\prime}_{\SOURCE{}i}$ and $\mcr_{\SOURCE{}i}$                                                       & L1Loss          \\\hline
\multirow{7}{*}{GBR} & Subsample          & Ratio of training sample                                                           & 0.8             \\\cline{2-4}
                     & Learning rate      & Coefficient shrinkage                                                     & \num{0.4}             \\\cline{2-4}
                         & Number of estimators & Max. number of gradient trees for model boosting          & \num{15}              \\\cline{2-4}
                     & Max. depth            & Maximum depth of a tree                                                       & \num{8}               \\\cline{2-4}
                     & Min. samples split    & Minimum number of samples for splitting the tree & \num{200}             \\\cline{2-4}
                     & Min. samples leaf     & Minimum number of samples for tree's leaves  & \num{40}              \\\cline{2-4}
                     & Loss                 & MAE between $\mcr^{\prime}_{\SOURCE{}i}$ and $\mcr_{\SOURCE{}i}$  & MAE\\
                     \hline
\end{tabular}}
\end{table}

\subsection{Evaluation metrics}
In this section, we evaluate the performance of LR, MLP, and GBR prediction models using five metrics.

\paragraph{Pearson correlation coefficient} between the microservice time and its call rate in the Alibaba trace:

\small
\begin{equation*}
    \frac{\sum\limits_{\SOURCE\in \mathcal{M}_{\SOURCE} \land m_i\in\mathcal{M}}\left(\mt_{i,j}-\overline{\mt_{i,j}}\right)\cdot\left(\mcr_{\SOURCE{}i}-\overline{\mcr_{\SOURCE{}i}}\right)}{\sqrt{\sum\limits_{\SOURCE\in \mathcal{M}_{\SOURCE} \land m_i\in\mathcal{M}}\left(\mt_{i,j}-\overline{\mt_{i,j}}\right)}\cdot\sqrt{\sum\limits_{\SOURCE\in \mathcal{M}_{\SOURCE} \land m_i\in\mathcal{M}}\left(\mcr_{\SOURCE{}i}-\overline{\mcr_{\SOURCE{}i}}\right)} }
\end{equation*}
\normalsize
where $\overline{\mt_{i,j}}$ and $\overline{\mcr_{\SOURCE{}i}}$ show the average microservice time and the call rate, respectively.

\paragraph{Predicted microservice call rate} $\mcr{^\prime}_{\SOURCE{}i}$ defined in Section~\ref{sec:predmodels}.

\paragraph{Number of replicas} $\mathcal{L}_{i}$ defined in Section~\ref{sec:obj}.

\paragraph{Mean absolute error} also referred to as L1Loss~\cite{L1Loss}, represents 
the average sum of absolute differences between the predicted microservice call rates $\mcr^{\prime}_{\SOURCE{}i}$ and $\mcr_{\SOURCE{}i}$, respectively, in the testing and training:
\begin{equation*}
    \mathrm {MAE} ={\frac{1}{\mathcal{N}_{\mathcal{M}}}}\cdot \sum\limits_{\SOURCE\in \mathcal{M}_{\SOURCE} \land m_i\in\mathcal{M}}\left|\mcr_{\SOURCE{}i}-\mcr^{\prime}_{\SOURCE{}i}\right|.
\end{equation*}

\paragraph{Mean absolute percentage error} ($\mathrm{MAPE}$) quantifies the prediction accuracy of an ML model:
\begin{equation*}
        \mathrm{MAPE} ={\frac{1}{\mathcal{N}_{\mathcal{M}}}\cdot\sum\limits_{\SOURCE\in \mathcal{M}_{\SOURCE} \land m_i\in\mathcal{M}}\left|\frac{\mcr_{\SOURCE{}i}-\mcr^{\prime}_{\SOURCE{}i}}{\mcr_{\SOURCE{}i}}\right|}.
\end{equation*}

\section{Experimental Results} \label{sec:experimntresult}
This section presents the performance evaluations of the ML models in predicting the microservice call rates and replicas.

\subsection{Feature distribution and correlation} 
\begin{figure}[!t]
    \centering
    \includegraphics[width=.9\columnwidth]{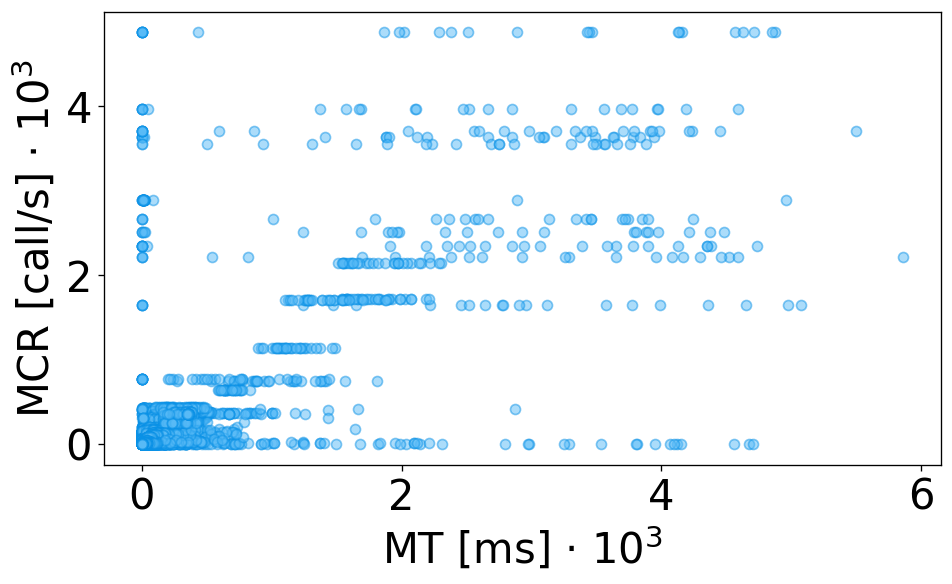}
    \caption{\centering Distribution and correlation of $\mt_{i,j}$ and $\mcr_{\SOURCE{}i}$ in Alibaba microservices dataset.}
    \label{fig:scatter}
\end{figure}

Figure~\ref{fig:scatter} shows the distribution, correlation, and relative variation of the two features $\mt_{i,j}$ and $\mcr_{\SOURCE{}i}$ in the Alibaba dataset using the Pearson coefficient. The results denote that we achieve a high correlation \SI{75}{\percent} between both feature sets, denoting that the prediction is applied to a correlated set of features.

\subsection{Model fitting}
\begin{figure}[t]
\centering
\subfloat[LR fitting.]{\includegraphics[width=\columnwidth]{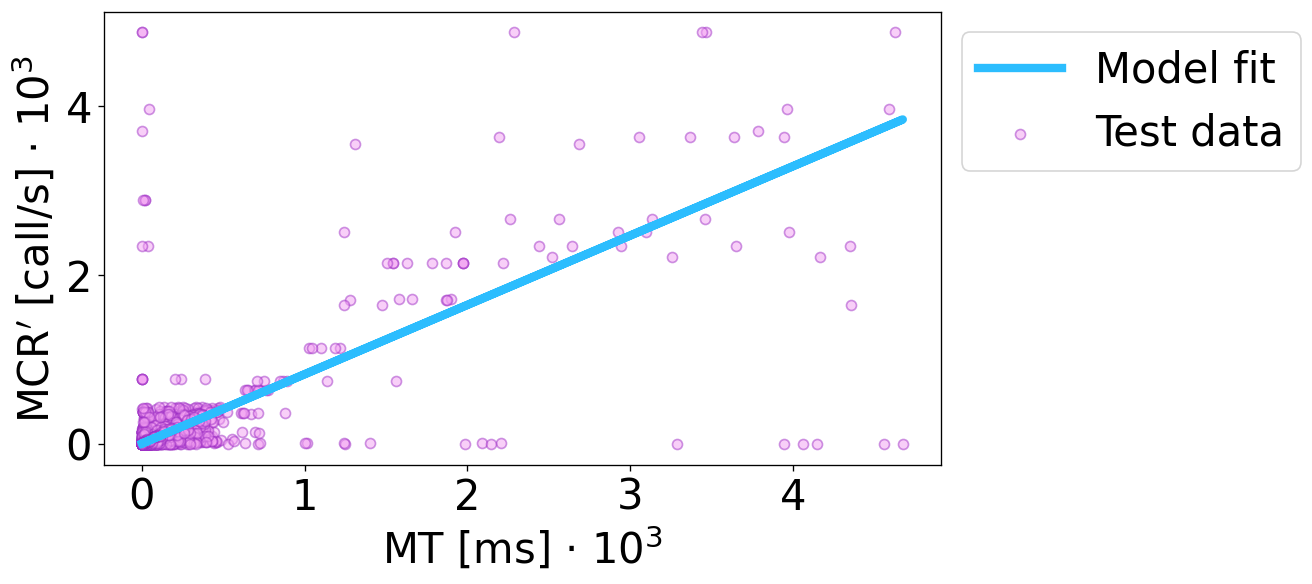}\label{fig:dist-lr}}\\
\subfloat[MLP fitting.]{\includegraphics[width=\columnwidth]{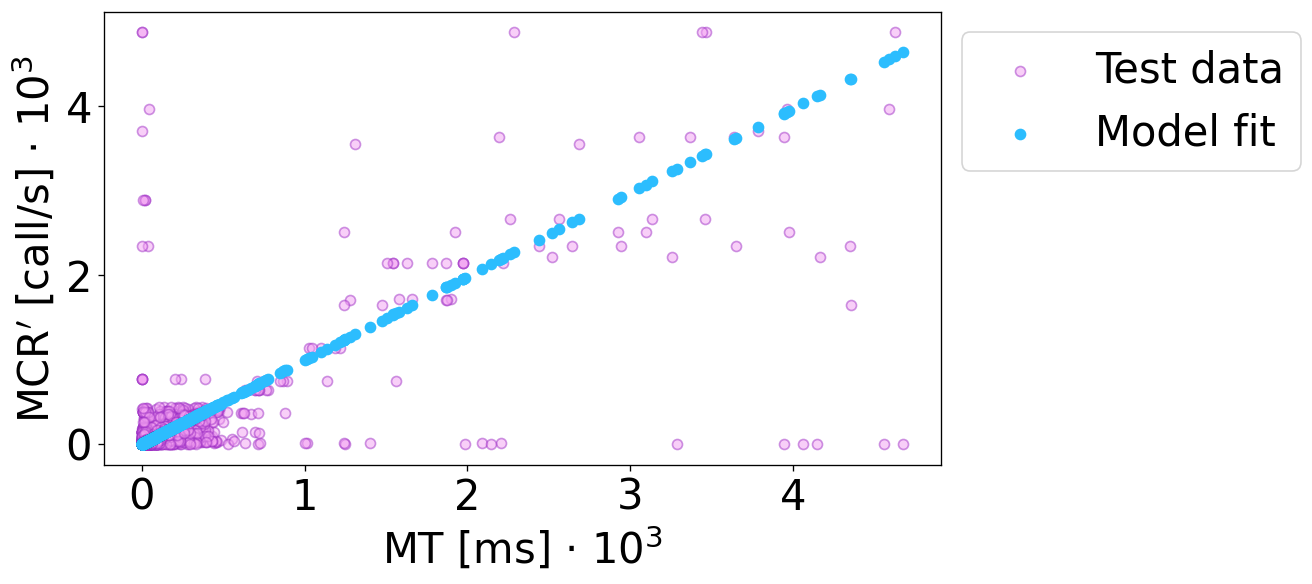}\label{fig:dist-mlp}}\\
\subfloat[GBR fitting.]{\includegraphics[width=\columnwidth]{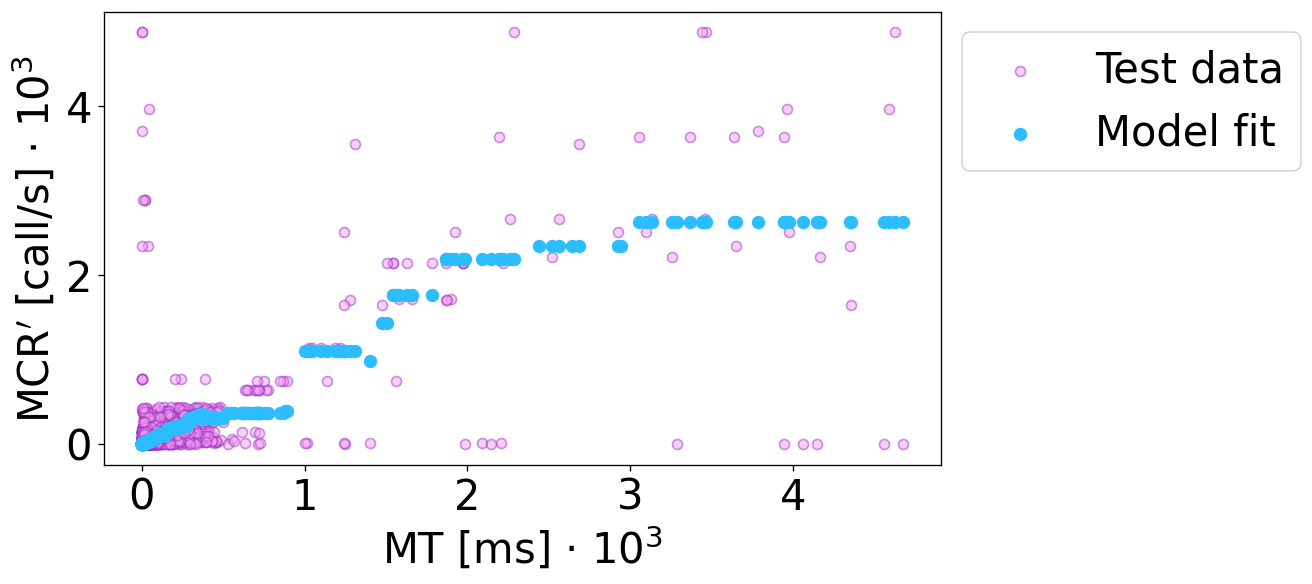}\label{fig:dist-gbr}}
\caption{ML model fitting to the test dataset.}
\label{fig:results}
\end{figure}

Figure~\ref{fig:dist-lr} shows that the LR model fits a linear relation between the predicted microservice call rate $\mcr^{\prime}_{\SOURCE{}i}$ and microservice time $\mt_{i,j}$. 
Figure~\ref{fig:dist-mlp} depicts fitting a linear MLP model to the test dataset. Although the model learns to fit a linear relation between the microservice time and call rate, it has a slightly lower MAE than the LR, as shown in Table~\ref{tab:traintime}. This indicates that the MLP model remains a  proper fit for this dataset despite its neural network baseline imposing a computationally intensive method compared to LR and GBR. 
Figure~\ref{fig:dist-gbr} shows that the GBR ensemble model does not follow a linear pattern because it iteratively fits new decision tree regressors to the loss of the previous ensemble. 
In other words, the model continuously tunes and boosts its predictions by fitting a new subset of training data to the ensemble of previous models to create a single low-error predictive model. 

\begin{table}[!t]
\centering
\caption{\centering MAE, MAPE, and training times of the prediction models.}
\label{tab:traintime}
\resizebox{\columnwidth}{!}{
\begin{tabular}{|c||c|c|c|}
\hline
\textit{Prediction model} & \textit{MAE} & \textit{MAPE} & \textit{Training time [\si{\second}]}\\
\hline
\hline
\textit{LR}&\numrange{10.25}{12.84}&\numrange{16.3}{16.4}&\numrange{1.3}{2}\\
\hline
\textit{MLP}&\SIrange{9.3}{9.4}{}&\numrange{16.5}{16.7}&\numrange{11}{14}\\
\hline
\textit{GBR}&\numrange{8.90}{8.94}&\numrange{11.5}{11.7}&\numrange{3}{4.5}\\
\hline
\end{tabular}}
\end{table}



\subsection{Training time} Table~\ref{tab:traintime} illustrates the superiority of the LR method that lowers the training time with the expense of increasing the prediction errors compared with the MLP and GBR. The neural network-based MLP increases the training time of the prediction, although it is based upon the linear models as defined in Section~\ref{sec:predmodels}. However, the GBR model reaches a balance between the prediction errors, including the MAE and MAPE, and the training time of the training model. 

\subsection{Number of replicas}  
Table~\ref{tab:replica} shows that the ML models estimate the number of replicas by following almost close prediction errors. The results show that the GBR model reaches lower MAPE regarding replication prediction compared to the LR and MLP.

\begin{table}[!t]
\centering
\caption{\centering MAPE of the prediction models.}
\label{tab:replica}
\begin{tabular}{|c||c|}
\hline
\textit{Prediction model} 
& \textit{MAPE}\\
\hline
\hline
\textit{LR}
&\numrange{14.3}{14.5}\\
\hline
\textit{MLP}
&\numrange{15.9}{16.2}\\
\hline
\textit{GBR}
&\numrange{10.9}{11.1}\\
\hline
\end{tabular}
\end{table}

\section{Conclusion and Future Work} \label{sec:conclusion}
We explored and compared three ML methods to improve the resource provisioning affected by stochastic changes due to the users' requirements by investigating the performance evaluation of a set of ML models on the monitoring data. 
We used three different ML models, LR, GBR, and MLP, that predict the microservice call rate based on the microservice time scheduled in the Alibaba Cloud resources. 
Since utilizing the MLP for this problem with one input and one output was complex, we set a small number of neurons and layers in its prediction model. 
The experimental results show that the GBR reduces the MAE and the MAPE compared to LR and MLP models. 
Moreover, the results show that the gradient boosting model estimates the number of replicas for each microservice close to the actual data without any prediction. In the future, we plan to explore integrating the ML models in the \texttt{Kubernetes} autoscaling component~\cite{horn2022multi} and evaluate the optimal deployment of microservices.

\section*{Acknowledgement} This work received partial funding from:
\begin{itemize}
\item \emph{European Union}'s grant agreements H2020 101016835 (DataCloud), HE  101093202 (Graph-Massivizer), and HE 101070284 (enRichMyData);
\item \emph{Austrian Research Promotion Agency (FFG)}, grant agreement 888098 (K{\"a}rntner Fog) .
\end{itemize}
\balance
\bibliography{ref}
\bibliographystyle{unsrt}
\end{document}